\documentclass{emulateapj}
\usepackage{cases}

\slugcomment{}

\shorttitle{Terrestrial planets across space and time}
\shortauthors{Zackrisson et al.}

\begin{document}
\title{Terrestrial planets across space and time}

\author{Erik Zackrisson\altaffilmark{1}$^*$, Per Calissendorff\altaffilmark{2}, Juan Gonz\'alez\altaffilmark{2}, \\Andrew Benson\altaffilmark{3}, Anders Johansen\altaffilmark{4}, Markus Janson\altaffilmark{2}}
\altaffiltext{*}{E-mail: erik.zackrisson@physics.uu.se}
\altaffiltext{1}{Department of Physics and Astronomy, Uppsala University, Box 515, SE-751 20 Uppsala, Sweden}
\altaffiltext{2}{Department of Astronomy, AlbaNova, Stockholm University, SE-106 91 Stockholm, Sweden}
\altaffiltext{3}{Carnegie Observatories, 813 Santa Barbara Street, Pasadena, CA 91101, USA}
\altaffiltext{4}{Lund Observatory, Department of Astronomy and Theoretical Physics, Lund University, Box 43, SE-221 00 Lund, Sweden}

\begin{abstract}
\noindent
The study of cosmology, galaxy formation and exoplanets has now advanced to a stage where a cosmic inventory of terrestrial planets may be attempted. By coupling semi-analytic models of galaxy formation to a recipe that relates the occurrence of planets to the mass and metallicity of their host stars, we trace the population of terrestrial planets around both solar-mass (FGK type) and lower-mass  (M dwarf) stars throughout all of cosmic history. We find that the mean age of terrestrial planets in the local Universe is $7\pm{}1$ Gyr for FGK hosts and $8\pm{}1$ Gyr for M dwarfs. We estimate that hot Jupiters have depleted the population of terrestrial planets around FGK stars by no more than $\approx 10\%$, and that only $\approx 10\%$ of the terrestrial planets at the current epoch are orbiting stars in a metallicity range for which such planets have yet to be confirmed. The typical terrestrial planet in the local Universe is located in a spheroid-dominated galaxy with a total stellar mass comparable to that of the Milky Way. When looking at the inventory of planets throughout the whole observable Universe, we argue for a total of $\approx 1\times 10^{19}$ and $\approx 5\times 10^{20}$ terrestrial planets around FGK and M stars, respectively. Due to light travel time effects, the terrestrial planets on our past light cone exhibit a mean age of just $1.7\pm 0.2$ Gyr. These results are discussed in the context of cosmic habitability, the Copernican principle and searches for extraterrestrial intelligence at cosmological distances.
\end{abstract}



\keywords{Planets and satellites: terrestrial planets -- galaxies: formation -- cosmology: miscellaneous -- extraterrestrial intelligence}


\section{Introduction} 
The use of transit photometry and radial velocity measurements have in recent years allowed the detection and characterization of large numbers of exoplanets in the same size and mass range as Earth (see \citealt{Winn & Fabrycky} for a recent review). By coupling the observed occurrence rate of such terrestrial planets to models of star and galaxy formation, it has now become possible to predict the prevalence of Earth-like planets in the Milky Way \citep[e.g.][]{Gonzalez01,Lineweaver01,von Bloh et al.,Lineweaver04,Prantzos,Gowanlock et al.,Guo et al.,Bonfils et al.} and Andromeda \citep{Carigi et al.,Spitoni et al.}, in other galaxies in the local volume \citep{Gonzalez01,Sundin,Suthar & McKay,Dayal15,Forgan15} and even throughout the observable Universe \citep{Wesson, Lineweaver07, Behroozi & Peeples,Olson,Dayal16,Loeb16}. 

The cosmic evolution of the planet population is relevant for a number of issues in the intersection between astrobiology, cosmology and SETI (the Search for Extraterrestrial Intelligence). The formation time and current age distribution of such planets enters arguments concerning the timing of biogenesis \citep[e.g.][]{von Bloh et al.,Cirkovic,Vukotic & Cirkovic}, the rate of planet-sterilizing events \citep{Tegmark & Bostrom} and anthropic selection biases affecting cosmological parameters \citep[e.g.][]{Lineweaver07,Egan & Lineweaver,Barreira & Avelino,Larsen et al.}, whereas the distributions of planets among galaxies of different type in the local Universe and throughout the observable Universe (i.e. galaxies on our past light cone) are relevant for the prospects of SETI on extragalactic scales \citep[e.g.][]{Annis,Wright et al. a,Wright et al. b,Griffith et al.,Olson,Zackrisson et al.,Garrett}. Here, we extend on previous attempts to model the cosmological distribution of planets by coupling semi-analytical models of galaxy formation to a recipe for planet formation that depends on both stellar mass and stellar metallicity. Using this machinery, we predict the spatial and temporal distribution of terrestrial planets (hereafter TPs) in both the local and distant Universe. Throughout this paper, we define TPs as planets in the size and mass range $\approx 0.5$--2.0 $R_\oplus$, $\approx 0.5$--10 $M_\oplus$ (i.e. both Earth-like planets and and Super-Earths), thereby including solid-surface planets up to largest sizes that could potentially allow habitable conditions according to the definition by \citet{Alibert}.

Our computational approach is similar to that of \citet{Behroozi & Peeples}, but differs in the details of the recipe for TP formation, by extending the inventory from solar-like stars (spectral type FGK) to also include low-mass stars (M dwarfs), in the treatment of the stellar metallicity distribution within each galaxy, by considering the impact of the lifetimes of FGK stars on the inventory of present-day TPs, and by presenting our census of TPs in the observable Universe in terms of galaxies on our past light cone (the case relevant for high-redshift SETI) rather than for the galaxies in the present-day Hubble volume. 

In section~\ref{models}, we describe our recipe for TP formation and the galaxy formation models used. Our predicted cosmic inventory of TPs in the local and distant Universe is presented in section~\ref{results}. Section~\ref{discussion} features a discussion on the potential habitability of these planets and the relevance of our results for the Copernican principle and extragalactic SETI. We also present a comparison to previous estimates to characterize the properties of the exoplanet population on cosmological scales and discuss the associated uncertainties. Section~\ref{summary} summarizes our findings.

\section{Models}
\label{models}
\subsection{Planet formation}
\label{planet_models}
Our recipe for TP formation is similar to that of \citet{Lineweaver01}, in which different metallicity-dependent probabilities are adopted for the formation of terrestrial and giant planets, and in which TPs are assumed to be destroyed by close-orbit giants (``hot Jupiters''). This computational machinery is here applied to a wider class of stars, with parameters updated to reflect recent advances in exoplanet studies and with planet formation probabilities that depend on both stellar metallicity and mass.   

\emph{Kepler} data show that the occurrence rate $f_\mathrm{TP}$ of TPs of size $R = 1$ -- $2\ R_\Earth$ with orbital periods $< 400$ days around solar type FGK-stars may be as high as $f_{\mathrm{TP},\mathrm{FGK}} \approx 0.40$ \citep{Petigura et al.}. Furthermore, the consensus of both transit and Doppler radial velocity surveys is that the occurrence rate of small rocky planets with $R = 0.5 - 2.0 R_\Earth$ orbiting M-dwarf stars is $f_{\mathrm{TP}, \mathrm{M}} \approx 1$ \citep[e.g.][]{DC13, DC15, Bonfils et al., Tuomi et al.}. 

Both observations and simulations indicate that stars in metal-enriched environments yield a greater number of giant planets compared to metal-poor environments \citep[e.g.][]{Armitage & Rice, Johnson et al.}. With increasing evidence for a metallicity correlation, \citet{Fischer & Valenti} proposed that the occurrence rate of close-in Jupiter-sized giant planets orbiting FGK-type stars can be described by a simple power law. Further investigations by e.g. \citet{Gaidos & Mann} suggests that the power law cannot be described by the same parameters for solar type stars (FGK) and low-mass M-dwarfs, implying that there is also a mass correlation embedded in the prevalence of giant planets. Here, we approximate the probability of forming giant planets $P_\mathrm{FG}$ as the giant planet occurrence rate $f_\mathrm{FG}$ described by \citet{Gaidos & Mann}:
\begin{equation}\label{eq:recipe}
P_{\mathrm{FG}}\approx f_{\mathrm{FG}}([\mathrm{Fe/H}], M_*) = f_0 10^{a [\mathrm{Fe/H}]} M_*^b,
\label{PFG_eq}
\end{equation}

where $M_*$ is the mass of the star in solar masses, $f_0$ a constant factor estimated to be 0.07 by \citet{Gaidos & Mann}, $a$ is the metallicity parameter estimated to be  $1.80 \pm 0.31$ and $1.06 \pm 0.42$ for spectral types FGK and M respectively and $b$ the mass correlation parameter assumed to be 1.

Following \citet{Lineweaver01}, we assess the probability of harboring TPs, $P_{\mathrm{HTP}}$, by estimating the number of TPs formed as described above, and appraise how many are destroyed by migrating giant planets described by Equation (~\ref{eq:recipe}) as

\begin{equation}
P_{\mathrm{HTP}} = P_{\mathrm{FTP}} (1 - P_{\mathrm{FG}}),
\label{PHTP_eq}
\end{equation}

where $P_{\mathrm{FTP}}$ is the probability of forming a TP and $P_{\mathrm{FG}}$ is the probability of forming a close-orbit giant planet, thus destroying the prospect of harboring a TP in the process. This approach admittedly neglects the intriguing possibility that TPs may reform in the wake of a migrating Jupiter or otherwise survive the ordeal \citep[e.g.][]{Fogg & Nelson}. Moreover, by adopting the parameter values for equation~(\ref{PFG_eq}) from \citet{Gaidos & Mann}, we are implicitly assuming that all close-in giant planets considered in their study (orbital period $< 2$ years) are detrimental for TPs -- including those that would be classified as warm rather than hot Jupiters. This likely overestimates the impact of close-in giants on the prevalence of TPs, since some warm Jupiters are known to have TP companions \citep{Steffen12,Huang16}. Because of these uncertainties, we in Section~\ref{results} present the fraction of TPs that is predicted to be lost due to close-in giants, so that the size of the population potentially missing from our inventory may be estimated.

It is very likely that the probability of forming TPs requires a minimum threshold of metallicity for the host star. At the low-metallicity end, we therefore assume a gradual decrease in the probability of forming terrestrial planets. In earlier work by e.g. \citet{Prantzos}, the probability to form Earth-like planets is envisioned as a step function going from zero probability to maximum probability at [Fe/H]$= - 1$. In our baseline model, we instead include stars with metallicities all the way down to [Fe/H]$\approx -2.2$, which corresponds to the level where the minimum mass solar nebula \citep{Hayashi81} would have insufficient solid materials to form a terrestrial planet ($M>0.5 M_\oplus$) in an Earth-like orbit. Since stars at such low metallicites would most likely struggle to form terrestrial planets, we adopt a $P_{FTP}$ function that includes a smooth cut-off $k(Z)$ at the low-metallicity end:
\begin{equation}
 P_{FTP}(Z) =  f_\mathrm{TP}\; k(Z).
\end{equation}
Here, we adopt $f_\mathrm{TP}=1$ for M dwarfs and $f_\mathrm{TP}=0.4$ for FGK stars and a low-metallicity cut-off given by: 
\begin{equation}
 k(Z) = \frac{Z - 0.0001}{0.001-0.0001} 
\end{equation}
for $-2.2 \leq$ [Fe/H] $\leq -1.2$, but $k(Z)=0$ for [Fe/H] $<-2.2$. 
We also assume that the probability for TP formation stays constant at [Fe/H] $>-1.2$ with $k(Z)=1$ at all higher metallicities. The function $k(Z)$ is admittedly arbitrarily chosen, but as we demonstrate in Section~\ref{local} and~\ref{Planet_formation_uncertainties}, the exact choice of cut-off function has an insignificant impact on our results.

To be consistent with the isochrones used to estimate stellar lifetimes (see Section \ref{galaxy_formation}), we adopt a solar metallicity of $Z_\odot=0.152$. 

\begin{figure}[t]
\centering
\includegraphics[width=84mm]{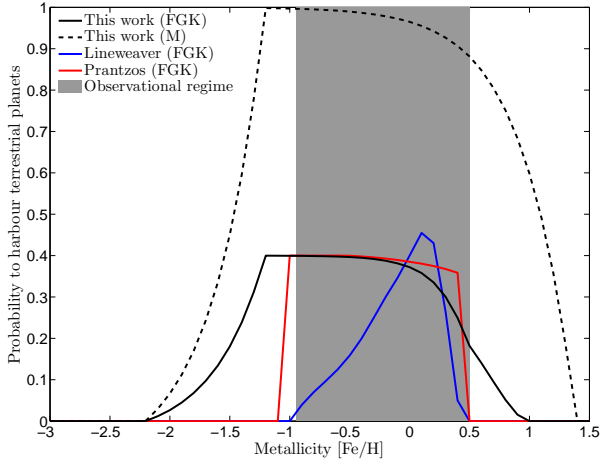}	
\caption{The adopted metallicity-dependent probability for FGK (black solid line) and M stars (black dashed) to harbor TPs. The blue line depicts the metallicity relation derived by \citet{Lineweaver01} and the red line the relation used by \citet{Prantzos}. The gray area describes the metallicity regime in which detections of TPs have so far been made.} 
\label{planet_recipe}
\end{figure}

The resulting probability for FGK and M stars to harbor TPs is shown in Figure~\ref{planet_recipe}. Our recipe implies an identical metallicity dependence at the low-metallicity end for FGK and M type stars, but a more pronounced probability decrease at the high-metallicity end in the case of FGK stars. This machinery for relating the formation and destruction probability for TPs is very similar to that used by \citet{Lineweaver01} and \citet{Prantzos} for FGK stars, but differs in the details of the adopted metallicity dependencies. As a result, our probability distribution for the occurrence of TPs around FGK stars is far less peaked around the solar value than that of \citet{Lineweaver01}, but still features a high-metallicity drop-off less pronounced than that of \citet{Prantzos}. 

Our TP model also differs in a number of ways compared to the recent works of \citet{Behroozi & Peeples} and \citet{Dayal15}, which do not assume that close-in giants have any adverse effects on the formation of TPs around FGK stars, or specifically treat the difference between FGK and M dwarfs. It should be stressed that in contrast to both \citet{Behroozi & Peeples} and \citet{Dayal15}, our $P_\mathrm{HTP}$ function in Equation.~\ref{PHTP_eq} offers no indication to whether TPs could be habitable or not. A discussion about the prospects of habitability is instead provided in Section~\ref{habitability}. \citet{Behroozi & Peeples}  adopt a minimum metallicity for TP occurrence of [Fe/H] $= -1.5$ based on \citet{Johnson & Li}, and a constant occurrence at higher metallicities, whereas \citet{Dayal15} explore cases where $P_\mathrm{HTP}$ either has no metallicity dependence, or a very strong one ($P_\mathrm{HTP}\propto Z$).

\subsection{Galaxy formation}
\label{galaxy_formation}
Semi-analytical models represents a powerful tool to connect high-redshift galaxies to their present-day descendants \citep[for a review, see][]{Baugh}. Here, we use such models to trace the formation of stars and the build-up of heavy elements within galaxies throughout cosmic history. 

In this paper, we base our primary results on the version of the Durham semi-analytic model GALFORM described in \citet{Baugh et al.} The Durham model has been demonstrated to reproduce the luminosity function of Lyman-break galaxies at $z\approx 3$, present-day optical and near- and far-infrared luminosity functions and the size-luminosity relation for $z\approx 0$ late-type galaxies \citep{Baugh et al.,Gonzalez et al. 09}. It is also in fair agreement with the observed cosmic star formation history \citep{Madau & Dickinson}. When possible, we also use the independent semi-analytic model Galacticus \citep{Benson} to cross-check the robustness of important conclusions.

From the Durham semi-analytical model, we extract the galaxy population within a simulated comoving box of side 64Mpc/h in 50 redshifts snapshots between z=0 and z=10. This gives us between $\approx 200$ ($z=10$) and $\approx 36000$ ($z=0$) galaxies per snapshot. The stellar mass range of these galaxies is $\sim 10^7$ to $\sim 10^{12}\ M_\odot$ at z=0. However, to avoid potential resolution problems at the low-mass end, we will throughout this paper focus on galaxies in the $M\gtrsim 10^8\ M_\odot$ mass range. This is not likely to affect our results in any significant ways, since both our own models and extrapolations of the observed galaxy stellar mass function to lower-mass objects \citep{Kelvin14,Moffett16} indicate that the fraction of stellar mass locked up in  $M<10^8\ M_\odot$ galaxies is at the sub-percent level.

From every redshift, we extract the internal age and metallicity distribution of the stars of all galaxies. By adopting the \citet{Kroupa01} universal stellar initial mass function (IMF), we derive the number of FGK type (assumed mass range 0.6--1.2  $M_\odot$) and M dwarf (0.08--0.6 $M_\odot$) subtypes in each metallicity and age bin. According to this IMF, $\approx 80\%$ of all stars in the 0.08--120 $M_\odot$ are born in the M dwarf mass regime whereas $\approx 10\%$ are born in the mass range of FGK stars. We then use mass- and metallicity-dependent main sequence lifetimes from the PARSEC v1.2S isochrones \citep{Bressan12,Chen14,Tang14} to remove dead stars from the inventory. 
 In these models, the lifetimes increase with metallicity, going from $\approx 5$ Gyr at $Z\approx 0.01\ Z_\odot$ to $\approx 12$ Gyr at $Z\approx 5\ Z_\odot$ for a mass with a zero age main sequence mass of 1.0 $M_\odot$. M dwarfs ($\leq 0.6\ M_\odot$) always have lifetimes longer than the present age of the Universe throughout this metallicity range and are therefore never removed. The adopted lifetimes are based on scaled solar elemental abundances, and while non-solar abundance ratios will affect the lifetimes, the enhancement of individual elements at the 0.3 dex level is not expected to affects the lifetimes of FG stars by more than $\sim 10\%$ \citep{Dotter07}.

Finally, we apply the mass- and metallicity dependent planet occurrence recipe of Section~\ref{planet_models} to calculate the number of TPs within the age and metallicity bin of each galaxy at every redshift.

\subsection{Cosmology}
\label{cosmology}
When converting redshifts into time and computing cosmological volumes required to reconstruct our past light cone, we adopt a flat $\Lambda$ Cold Dark Matter ($\Lambda$CDM) cosmology with $\Omega_{M}=0.308$, $\Omega_{\lambda}=0.692$ and Hubble constant $H_0 = 67.8$ km s$^{-1}$ Mpc$^{-1}$ \citep{Planck15}, giving a current age to the Universe of $t_0\approx 13.8$ Gyr. When comparing ages of typical planetary ages at various cosmological epochs to the age of the Earth, Earth is assumed to be 4.54 Gyr old \citep{Dalrymple}. 

\section{Results}
\label{results}

\subsection{Terrestrial planets in the local Universe}
\label{local}
In Table~\ref{statistics_local}, we present a number of statistical quantities predicted by our model for the population of TPs in the local Universe (i.e. averaged over all galaxies within the simulated volume at redshift $z=0$). According to our model, the global number of TPs per stellar population mass has a current average of $\approx 1.4\ M_\odot^{-1}$ at redshift $z=0$. Since M dwarfs greatly outnumber the FGK stars, this quantity is completely dominated by TPs around M dwarfs, and the corresponding value for TPs around FGK hosts is about 30 times lower ($\approx 0.045 \ M_\odot^{-1}$).  Close-in giants are expected to have diminished the current population of TPs around FGK stars by no more than $\approx 11\%$. Given our assumptions in Section~\ref{planet_models}, this fraction is even smaller for M dwarfs ($\approx 2\%$). By $z=0$, the exhausted lifetimes of FGK host stars have reduced the TP population around such stars by $\approx 1/3$, although this fraction varies slightly with galaxy mass. 

\subsubsection{Age distribution at $z=0$}
As shown in Table~\ref{statistics_local}, the average age of TPs around FGK stars and M dwarfs is currently $\approx 7.3$ Gyr and $\approx 8.4$ Gyr, respectively (i.e. 2.8 and 3.9 Gyr older than Earth, respectively). The oldest TPs in our simulation have ages of around $\approx 13$ Gyr (by comparison, the oldest TP known so far has an age of $\approx 11$ Gyr; \citealt{Campante+15}). The reason for the slightly lower ages of FGK host planets is mainly due to the shorter lifetimes of such stars. However, the fraction of FGK planets lost due to the exhausted lifetimes of their host stars is no more than 35\%, since only the most massive stars (mass higher than $\approx 0.95\ M_\odot$, although with a slight metallicity dependence) have main sequence lifetimes shorter than the current age of the Universe.

\begin{table}[t]
{\scriptsize
\caption{Terrestrial planets in the local Universe}
\begin{tabular}{lllll}
\hline
& Host & & &\\
& FGK & M & FGKM &\\
\hline
Mean age of terrestrial planet (Gyr) & 7.3 & 8.4 & 8.3 &\\
Terrestrial planets per $M_\odot$ & 0.045 & 1.3 & 1.4 &\\
Fraction lost to close-in giants & 0.11 & 0.02 & 0.02 &\\
Fraction outside observed [Fe/H] range & 0.04 & 0.1 & 0.09 &\\
\hline
\label{statistics_local}
\end{tabular}\\
}
\end{table}

\begin{figure}[t]
\centering
\includegraphics[width=84mm]{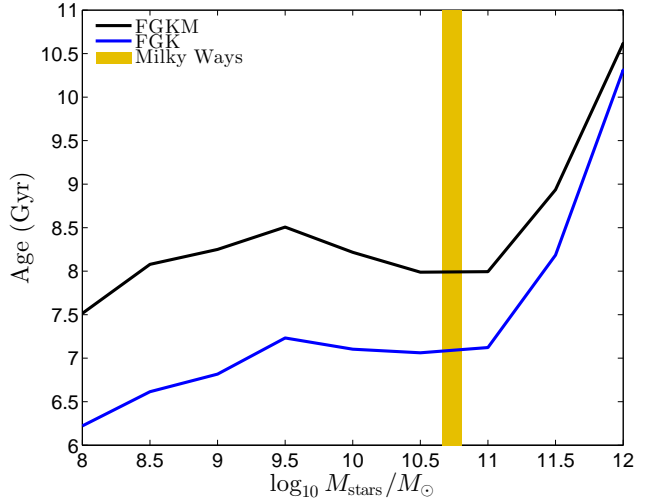}
\caption{The mean age of TPs as a function of galaxy stellar mass at $z=0$. The black line represents TPs around FGKM stars, whereas the blue line represents the FGK subpopulation. The amber-colored patch indicates the stellar mass of the Milky Way.} 
\label{age_vs_mass_lowz}
\end{figure}

The mean age of TPs shows a dependence on current galaxy mass, as shown in Figure~\ref{age_vs_mass_lowz}. The highest-mass galaxies (giant elliptical galaxies) harbor the oldest planets, as expected from the high average ages of stars in such systems. While the age difference between TPs around FGK and M stars in these galaxies is insignificant, TPs around FGK stars become slightly younger than those around M stars in lower-mass galaxies, and this offset reaches a maximum value of 1.2 Gyr at the low-mass end. This trend is due to a complicated interplay between the mass-metallicity relation of galaxies, the metallicity-lifetime relation for stars and the different star formation histories of low- and high-mass galaxies. 

Galaxies at the high-mass end of Figure~\ref{age_vs_mass_lowz} have higher metallicities, which leads to FGK stars with longer lifetimes. On the other hand, the stars in the highest-mass systems are also typically older than in the lowest-mass galaxies. This leads to a similar fraction of TPs around FGK stars being lost due to exhausted stellar lifetimes in $\sim 10^8\ M_\odot$ and $\sim 10^{12}\ M_\odot$ systems. However, the effect on the mean age of stars surviving until $z=0$ is different in the two cases, since the $\sim 10^8\ M_\odot$ galaxy populations have far more varied star formation histories and consequently a much broader age distribution. In the $\sim 10^{12}\ M_\odot$ galaxies, the removal of the very oldest FGK stars has an insignificant impact on the mean age, since most stars already lie in a limited age interval close to the mean. 

\subsubsection{Metallicity distribution at $z=0$}

\begin{figure}[t]
	\centering
	\includegraphics[width=84mm]{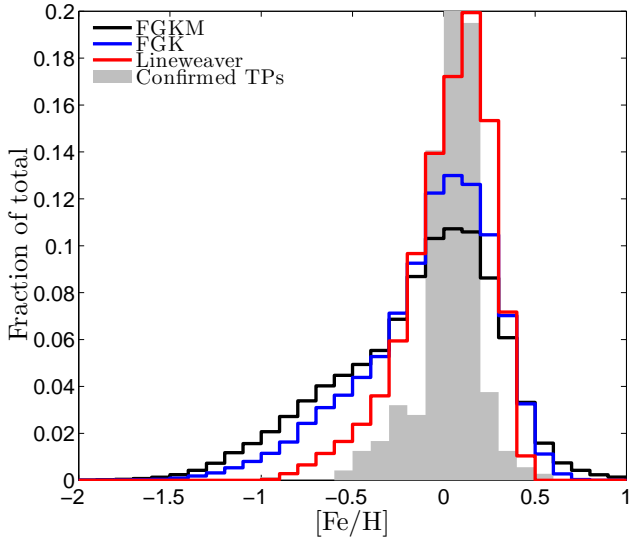}
	\caption{Normalized metallicity distribution for TPs in the galaxy population at $z=0$. The black and blue solid lines represent the predictions for FGKM and FGK stellar hosts in the case of our default metallicity dependence for TP formation. The red line depicts the corresponding case for FGK stars under the assumption of the \citet{Lineweaver01} metallicity dependence. The gray region indicates the much narrower metallicity distribution of currently confirmed TPs \citep[][]{Han14}, which due to the metallicity distribution within the Milky Way disk is much more strongly peaked than our prediction for the cosmological distribution. For clarity, the plotting range is limited to a fraction of 0.2, even though the largest bin of the observed distribution actually reaches 0.5.}
	\label{Zdist_lowz}
\end{figure}

The full metallicity distribution of the TP population in the local Universe is shown in Figure~\ref{Zdist_lowz}. The distribution for FGK hosts (blue), is slightly narrower than that for FGKM hosts (black; dominated by M dwarfs) due to two separate mechanisms that preferentially remove FGK stars in the low- and high-metallicity tails, respectively. Low-metallicity FGK planets have shorter lifetimes than high-metallicity ones, which means that a larger fraction of TP stars are lost at the low-metallicity end of the distributions. TPs around high-metallicity stars are on the other hand more efficiently destroyed by migrating giants, which removes TPs from the high-metallicity end as well. 

The red line demonstrates the effect of using the original metallicity relation by \citet{Lineweaver01}, which makes the distribution peak more strongly around the solar metallicity. For comparison, we also plot the distribution of stellar metallicities for currently confirmed TPs \citep[Data from \url{exoplanets.org};][]{Han14}, which is heavily biased by the selection function of stars in the Milky Way disk and consequently much narrower than our predictions for the cosmological distribution of TPs.

Our model predicts that at redshift $z=0$, $\lesssim 10\%$ of the TPs are orbiting FGK and M type stars in the metallicity range for which TPs have yet to be detected ([Fe/H]$< -0.95$ and [Fe/H]$>0.5$; \citealt{Morton16}). As further discussed in Section~\ref{Planet_formation_uncertainties}, this conclusion is remarkably robust to the details of the adopted metallicity dependence of the planet occurrence model at [Fe/H]$<-1.2$. Even if the probability to harbor TPs were to remain constant down to zero metallicity, this would boost the total number of TPs outside the metallicity range of current detections by less than 1\%, simply because the fraction of stars at these very low metallicities is so small.

\subsubsection{Terrestrial planets as a function of galaxy mass at $z=0$}
\begin{figure}[t]
\centering
\includegraphics[width=84mm]{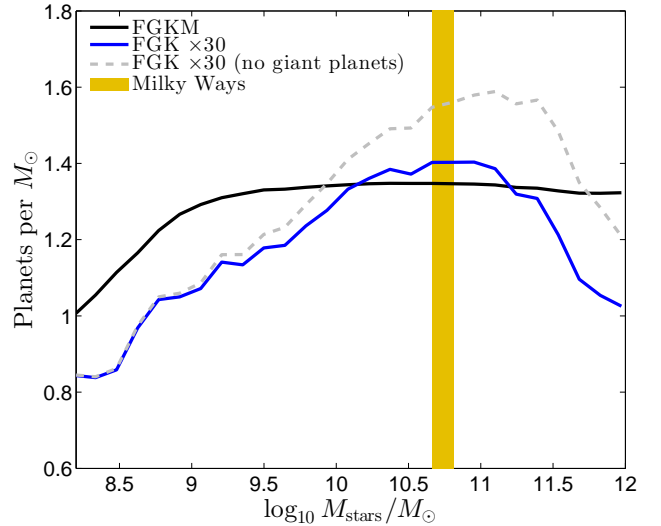}
\caption{TPs per unit stellar population mass for galaxies of different mass at $z=0$. The relation for FGK hosts (blue line) has been boosted by a factor of $30$ to illustrate the difference in shape compared to the relation for FGKM hosts (black line). The amber-colored patch indicates the mass of the Milky Way, the dashed gray line the predictions for FGK planets when the effect of close-in giants is turned off.}
\label{TP_per_mass}
\end{figure}

As seen in Figure~\ref{Zdist_lowz}, our predicted distribution has more planets at [Fe/H] $\leq -0.25$ than predicted by the \citet{Lineweaver01} metallicity relations. This directly affects the expected prevalence of TPs in low-mass galaxies, where such metallicities are common. The \citet{Lineweaver01} recipe would render galaxies at $M_\mathrm{stars}\lesssim 10^9\ M_\odot$ largely barren of TPs. No corresponding decline in planet numbers is predicted by our model, as demonstrated in Figure~\ref{TP_per_mass}, where the predicted number of TPs per stellar population mass is plotted as a function of galaxy mass. 

Instead, this quantity stays roughly constant at $\approx 1.4$ TPs per stellar population mass for TPs around FGKM host stars throughout the $M_\mathrm{stars}=10^9$--$10^{11}\ M_\odot$ galaxy mass range, and only drops by $\approx 30$\% for FGKM host stars in low-mass galaxies ($\sim 10^{8}\ M_\odot$) compared to galaxies in the mass range of the Milky Way (stellar mass 4--6$\times 10^{10}\ M_\odot$; \citealt{McMillan}). This reduction in TPs per total stellar mass in low-mass galaxies is due to a lack of heavy elements from which TPs are made, and is further augmented in the case of FGK host stars due to the reduced lifetimes of such stars at low metallicities. The number of TPs around FGK stars also drops at high metallicities, primarily because of the high average ages of stars in the most massive galaxies. These systems also have the highest metallicities, but the boosted stellar lifetimes resulting from this are not quite sufficient to counterweight the exhausted-lifetimes effect. The gray dashed line shows the corresponding predictions with the effect of close-in giants on TPs around FGK stars switched off. As seen, close-in giants have also reduced the planet population in high-mass galaxies, but this effect affects a much wider range of galaxy masses and is not responsible for the downturn in the the number of TPs around FGK stars per stellar population mass at $M_\mathrm{stars}\gtrsim 3\times 10^{11}\ M_\odot$.

A typical galaxy in the mass range of the Milky Way (stellar mass 4--6$\times 10^{10}\ M_\odot$; \citealt{McMillan}) is expected to harbor 5--$8\times 10^{10}$ TPs around M stars and 2--$3\times 10^9$ TPs around FGK stars. While our simulated galaxy catalogs are inadequate for addressing planet formation for galaxies $M<10^{8}\ M_\odot$, the observed-mass metallicity relation for dwarf galaxies \citep{Kirby13} suggests that the low-metallicity cut-off in planet formation adopted in this work would not cause any order-of-magnitude decline in the planet population of galaxies until one enters the $\lesssim 10^7 \ M_\odot$ galaxy mass range.

Another interesting consequence of Figure~\ref{TP_per_mass} is that galaxies like the Large Magellanic Cloud (LMC; stellar mass $M_\mathrm{stars}\approx 2\times 10^9\ M_\odot$; \citealt{Harris & Zaritsky}) are predicted not to be significantly depleted in TPs, but to harbor $\approx 7\times 10^7$ (around FGK stars) and $\approx 3\times 10^9$ (FGKM stars) of these. This is stark contrast to the conclusion reached by \citet{Zinnecker}, who argued that galaxies in the LMC metallicity range may be unable to form TPs as large as Earth. However, we note that transit data have already started to turn up rocky planets even larger than Earth \citep[][]{Buchhave+12,Torres+15} at metallicities at the upper range of those relevant for the inner regions of LMC \citep{Piatti & Geisler}. Based on the observed age-metallicity relation for the LMC \citep{Piatti & Geisler} and the LMC star formation history of \citet{Harris & Zaritsky}, our metallicity recipe for planet occurrence suggests that the average TP in the LMC should be significantly younger than the cosmic average -- the mean age is predicted to be about 1 Gyr older than Earth for TPs around M dwarfs, and 0.8 Gyr younger than Earth for TPs around FGK stars.

\begin{figure}[t]
\centering
\includegraphics[width=84mm]{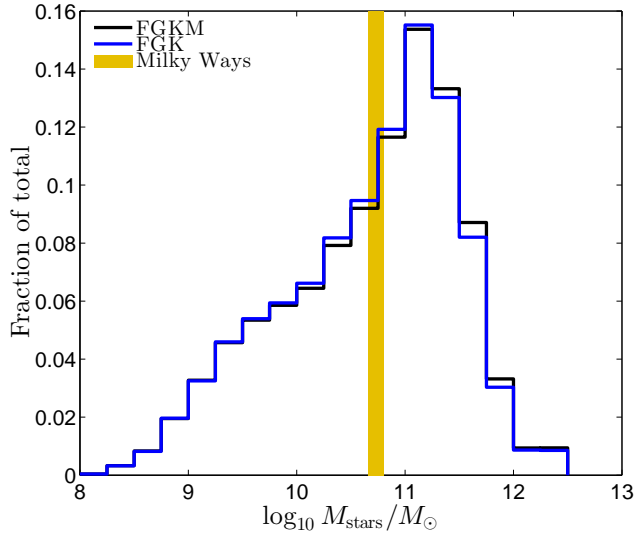}
\caption{The distribution of TPs across the $z=0$ galaxy population. The differently colored lines in the two panels represent the FGKM (black solid) and FGK (blue solid) hosts. The typical (median) TP is sitting in a galaxy with stellar mass $M_\mathrm{stars}\approx 6\times 10^{10}\ M_\odot$. The amber-colored patch indicates the mass of the Milky Way.}
\label{Planetdist_lowz}
\end{figure}

In Figure~\ref{Planetdist_lowz}, we show the distribution of TPs across the galaxies of different masses at $z=0$. The median galaxy mass of the TP-weighted distribution is $\approx 6\times 10^{10}\ M_\odot$, i.e. quite similar to that of the Milky Way (wheres the mean is $\approx 2\times 10^{11}\ M_\odot$). Interestingly, this suggests that most TPs are locked up in spheroid-dominated galaxies and not in a disk-dominated galaxy like the Milky Way, since the former type starts to dominate at stellar masses $M\gtrsim 10^{10}\ M_\odot$ \citep{Kelvin14,Moffett16}. The fact that Earth does not appear to be located in the most common type of TP-bearing galaxy, and the how this relates to the Copernican/mediocrity principle, is further discussed in Section~\ref{mediocrity_principle}.

\subsection{Terrestrial planets throughout the observable Universe}
\label{highz}
In our census of TPs throughout the observable Universe, we consider the population of TPs within all galaxies on our past light cone -- i.e. within a spherical volume whose outer radius is bounded by the maximum light travel time distance allowed by the current age of the Universe, and within which the cosmic epoch considered is a function of distance (or, equivalently, redshift $z$) from us. Galaxies, stars and planets at larger distances within this volume are seen at progressively earlier epochs in the history of the Universe. This definition of ``observable Universe'' differs from that of \citet{Behroozi & Peeples}, which instead consider the {\it current} state of the planet population in a cosmological volume of this order. However, due to the finite speed of light, the latter population cannot be directly observed.

In Table~\ref{statistics_highz} we present our results concerning the statistical quantities for the population of TPs on our past light cone. The total number of TPs around FGKM stars in the observable Universe is estimated to be $\approx 5\times 10^{20}$. Analogous to the case at $z\approx 0$ (Section~\ref{local}), this number is completely dominated by M dwarf planets ($\approx 98\%$), and the corresponding number for FGK stars is $\approx 1\times 10^{19}$. 

\begin{figure}[t]
	\centering
\includegraphics[width=84mm]{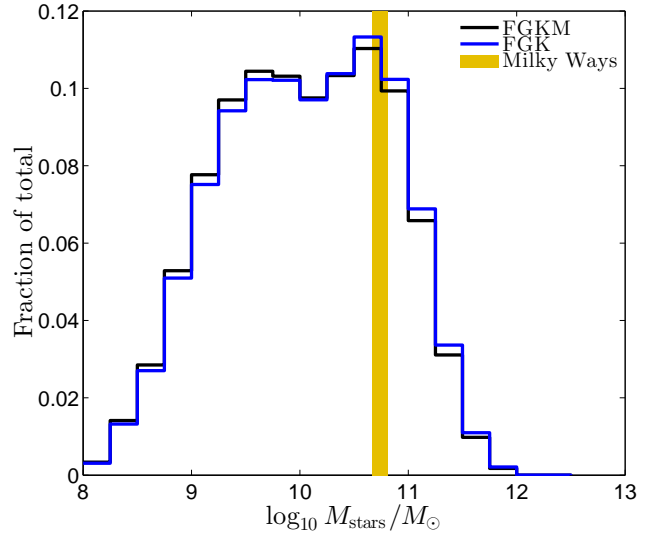}
\caption{The distribution of TPs across galaxies of different stellar masses on our past light cone. The differently colored lines represent the FGKM (black solid) and FGK (blue solid) stellar hosts. The typical (median) TP is sitting in a galaxy with stellar mass $M_\mathrm{stars}\approx 1\times 10^{10}\ M_\odot$. The amber-colored patch indicates the mass of the Milky Way.}
\label{Planetdist_highz}
\end{figure}

\subsubsection{Age distribution at high redshift}
\begin{figure*}[t]
\centering
\includegraphics[width=84mm]{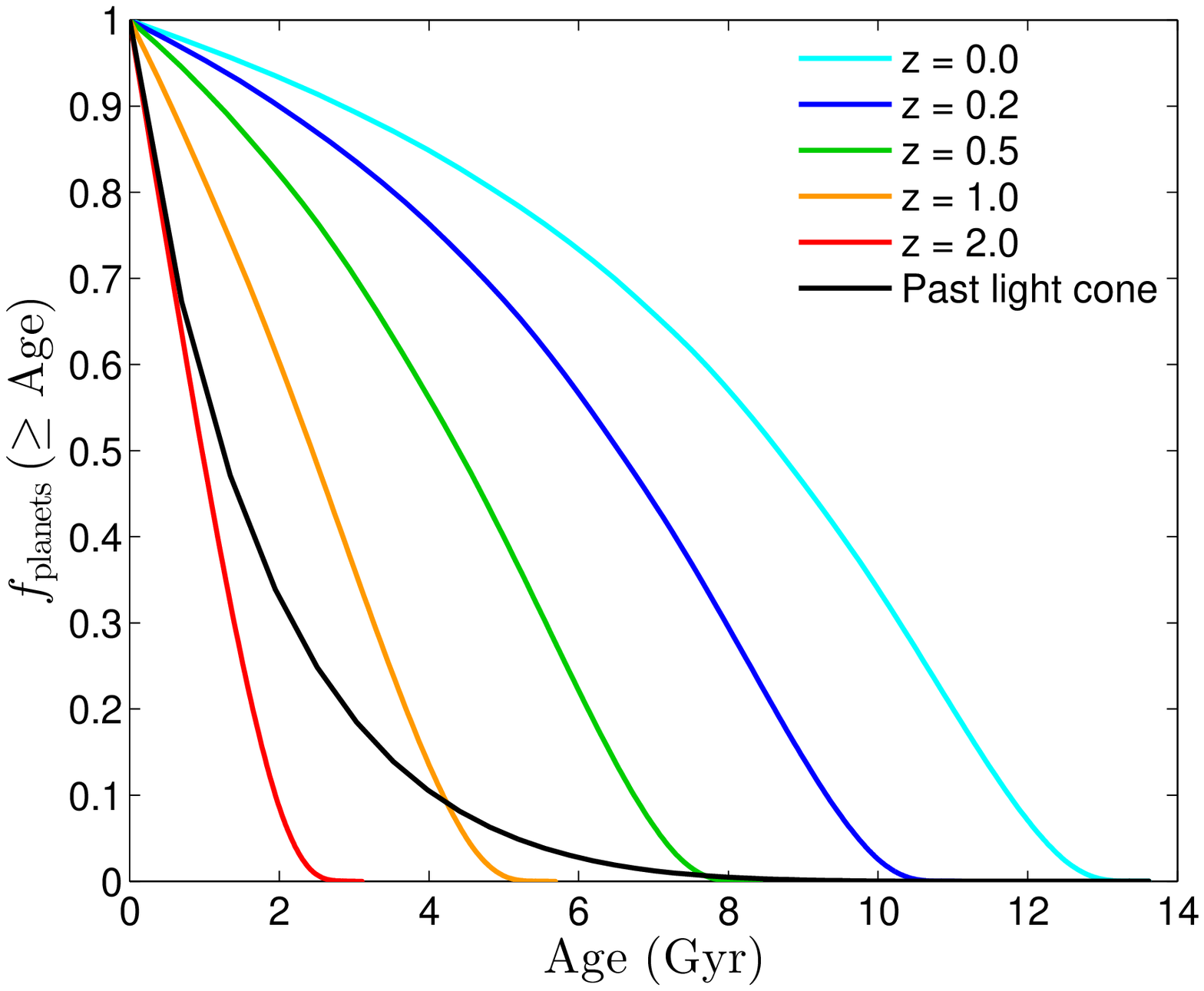}	
\includegraphics[width=84mm]{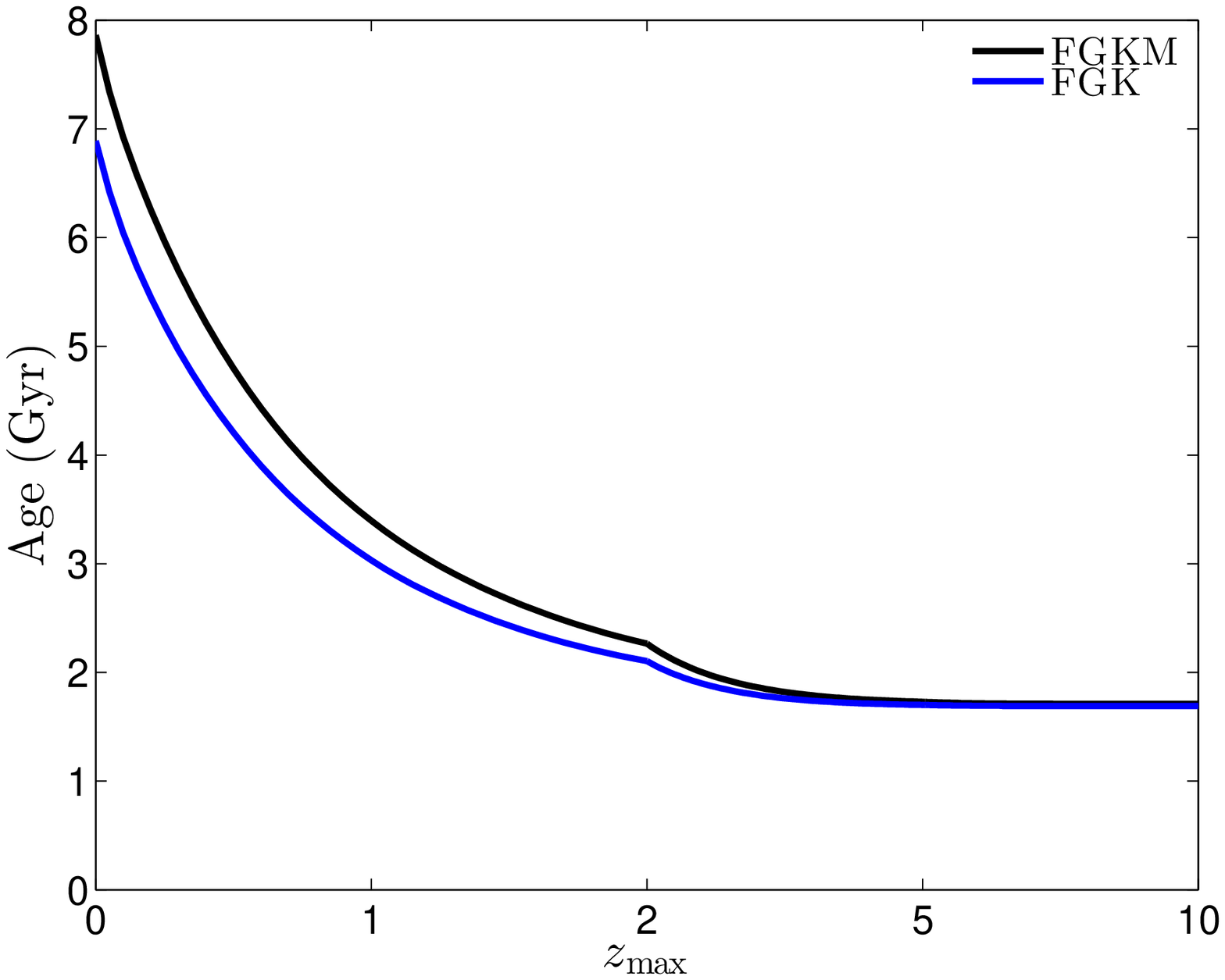}
\caption{TP age distribution as a function of redshift. {\bf a)} Age distribution of TPs at various discrete redshifts, including the distribution predicted when the entire observable Universe (all objects on our past light cone) is considered. {\bf b)} The mean age of TPs around FGK (blue line) and M-type hosts (black) within the volume on our past light cone out to limiting redshift $z_\mathrm{max}$. The cosmic mean of 1.7 Gyr is reached at $z\approx 5$.}
\label{agedist_highz}
\end{figure*}

Since the typical ages of TPs decrease as more distant regions on our past light cone are considered, the average TP age in the observable Universe comes out at $\approx 1.7$ billion year, i.e. almost 3 Gyr {\it younger} than Earth. The age distribution at a few different redshifts, along with the total distribution for the whole observable Universe is shown in Figure~\ref{agedist_highz}a. In Figure~\ref{agedist_highz}b, we show how the mean TP age for all objects on the past light cone evolves as the maximum considered redshift is increased. This mean TP age approaches the average age of TPs throughout the whole observable Universe reported in Table~\ref{statistics_highz} at $z\approx 5$ and changes insignificantly as higher redshifts are considered, due to the small number of planets at earlier epochs (as further discussed in Section~\ref{light_cone_dist}). The difference in mean age between FGK and M-type host planets also becomes increasingly insignificant at larger limiting redshifts, since fewer FGK stars in the high redshift have died because of old age. 

\subsubsection{Terrestrial planets as a function of galaxy mass at high redshift}
Since galaxy formation progresses hierarchically, high-mass galaxies were more rare in the past. The mean galaxy mass was therefore lower at earlier epochs on our past light cone, and the distribution of TPs across the galaxy population is therefore somewhat different compared to the local Universe. This is demonstrated in Figure~\ref{Planetdist_highz}, which shows that the typical TP on our past light cone is sitting inside a galaxy with a typical mass considerably lower than the that at $z=0$ (Figure~\ref{Planetdist_lowz}). The median galaxy mass in the distribution in Figure~\ref{Planetdist_highz} is $\approx 1\times 10^{10}\ M_\odot$, whereas the mean galaxy mass is $\approx 4\times 10^{10}\ M_\odot$.

\subsubsection{Distribution on the past light cone}
\label{light_cone_dist}
{The small contribution from the highest-redshift TPs to the statistics for the whole observable Universe can be understood from Figure~\ref{planetsinsky}, where we plot the number distribution of TPs as a function of redshift within the volume spanned by our observable Universe. The modest contribution from TPs at the highest redshift is mainly due to inefficient star formation (and not inefficient TP formation) at these early epochs. Even though the light cone volume increases by a factor of $\approx 2$ between $z=5$ and $z=10$, this is offset by the lower star formation rate density and the smaller time interval per unit redshift in the early history of the Universe. Consequently, stars at $z>5$ only add a few percent to the total stellar mass on our past light cone.}

In Figure~\ref{planetsinsky}, we also show the cumulative number of TP within the volume subtended by our past light cone out to a redshift $z_\mathrm{max}$. As this redshift limit approaches that of the earliest epochs of star formation, the cumulative number of TPs approaches the total number of  TPs in the whole observable Universe. As seen, the number increases relatively little beyond the peak of cosmic star formation \citep[$z\approx 2$;][]{Madau & Dickinson}, and we consequently find that 90\% of all TPs on our past light cone are at $z<3.3$. When considering only planets older than Earth, this limit is reached at even lower redshifts, since no objects older than Earth can exist at $z>1.4$ in the adopted cosmology. Hence, we find that 90\% of the TPs older than Earth are at $z<0.8$.

\section{Discussion}
\label{discussion}

\subsection{Habitability}
\label{habitability}
While the focus of this paper is on the prevalence of {\it TPs} on cosmological scales, it is primarily the subset of TPs that are able to sustain life that are relevant for astrobiology and SETI. While the requirements for planet habitability remains much debated \citep[for a recent review see][]{Gonzalez14}, the most common approach when discussing habitability on galactic scales is to start from a condition on the ability for planets to maintain liquid water (usually based on some variation on the moist greenhouse circumstellar zone of \citealt{Kasting93}; e.g. \citealt{Kopparapu13,Petigura et al.}). 

Current attempts to estimate the occurrence rate $f_{\mathrm{HTP}}$ of TPs located in the circumstellar habitable zone indicate a fraction $f_{\mathrm{HTP}}\approx 0.1$ for solar-type stars \citep{Petigura et al.,Batalha14} and $f_{\mathrm{HTP}}\approx 0.3$--0.5 for M dwarfs \citep{Kopparapu et al.,DC15}. These estimates are lower by factors of $\approx 1/4$ (FGK hosts) and $\approx 1/3$--1/2 (M dwarfs) compared to the occurrence $f_\mathrm{TP}$ rates adopted in section~\ref{planet_models} ($f_{\mathrm{TP},\mathrm{FGK}} \approx 0.40$; $f_{\mathrm{TP}, \mathrm{M}} \approx 1$) for the full TP population. Hence, a first-order approach would be to simply scale down the number of planets in Table \ref{statistics_local} and \ref{statistics_highz} by factors of this order.

\begin{table}[t]
{\scriptsize
\caption{Terrestrial planets in the observable Universe}
\begin{tabular}{lllll}
\hline
& Host & & &\\
& FGK & M & FGKM &\\
\hline
Total number & 1.1e19 & 4.7e20 & 4.8e20 &\\
Mean age of terrestrial planet (Gyr) & 1.7 & 1.7 & 1.7 &\\
Terrestrial planets per $M_\odot$ & 0.035 & 1.3 & 1.3 &\\
Fraction lost to close-in giants & 0.11 & 0.01 & 0.02 &\\
Fraction outside observed [Fe/H] range & 0.06 & 0.15 & 0.15 &\\
\hline
\label{statistics_highz}
\end{tabular}\\
}
\end{table}

So far, we have included all planets of size $R = 1.0 - 2.0 R_\Earth$ in our definition of TPs. However, the habitability of super-Earths ($R = 1.5 - 2.0$) is still a matter of debate. \citet{Seager et al.} argue that super-Earths need to be studied in individual cases, as those that develop a gaseous envelope may not be very habitable. \citet{Adibekyan et al.} also argue for a metallicity dependency for the potential habitability of super-Earths, arguing that super-Earths around metal-poor stars are found on tighter orbits than super-Earths around more metal-rich stars. By removing super-Earths from our estimates (and hence only include terrestrial planets of size $R = 1.0 - 1.5 R_\Earth$) using the relative distribution of sizes from \citet{Petigura et al.} for FGK hosts and \citet{DC15} for M-dwarf hosts, we reduce the occurrence rates of TPs in habitable zones to $f_{\mathrm{HTP},\mathrm{FGK}}\approx 0.04$ and $f_{\mathrm{HTP},\mathrm{M}}\approx  0.16$, i.e. a reduction by a factor of $\approx 1/2$ to the corresponding estimates including super-Earths.

Another open question is whether M-dwarfs can be associated with habitability in the same sense as solar-type stars. The inward migration of the habitable zone during the early stages of evolution of M dwarfs, along with the X-rays, extreme ultraviolet radiation and flares produced by these stars, may drastically erode a stable atmosphere and the prospects of liquid water \citep[e.g.][]{Segura05,Scalo07,Ramirez14,Luger15,Tian15}. \citet{Sengupta} argue that only a handful of the M-dwarfs with confirmed planets in the habitable zone meet the criteria for having Earth-like habitable conditions. Planets within the habitable zone of low-mass stars may also experience synchronous rotation and be tidally locked \citep{Kasting93}, leading to an unstable climate \citep{Kite+11}. 

Hence, a most conservative approach could involve excluding M dwarfs from the habitability discussion altogether and simply consider the FGK planets in Table \ref{statistics_local} and \ref{statistics_highz}. Adopting this strategy and removing super-Earths from the TP inventory would then scale down these estimates by a factor $f_{\mathrm{HTP},\mathrm{FGK}}/f_{\mathrm{TP},\mathrm{FGK}}\approx 0.04/0.4 = 0.1$, giving $\sim 5\times 10^{-3}$ habitable planets around FGK stars per unit total stellar population mass ($M_\odot$) and $\sim 1\times 10^{18}$ habitable planets around FGK stars in the observable Universe.

On galactic scales, there are possibly mechanisms other than the circumstellar habitable zone that affects the ability of planets to sustain life. Proposed effects include supernovae \citep[e.g.][]{Ruderman,Lineweaver04,Prantzos,Gowanlock et al.,Carigi et al.,Spitoni et al.,Dayal15}, gamma-ray bursts \citep[e.g.][]{Thomas et al.,Ejzak et al.,Piran & Jimenez,Dayal16}, active galactic nuclei \citep{Clarke81,Gonzalez05,Dayal16}, cosmic rays \citep[e.g.][]{Atri et al.}, Oort cloud comet perturbations \citep[e.g.][]{Feng14}, encounters with interstellar clouds \citep{Kataoka} and the dynamical and annihilation effects of dark matter \citep{Rampino}. \citet{Mellot & Thomas} presents an excellent review on various types of radiation-based threats to life.

While mechanisms of this type are outside the scope of this paper, the expected redshift trend (which is relevant for the SETI argument in section~\ref{seti}) is to make life on TPs more difficult to sustain at earlier times in the history of the Universe \citep[see also][]{Dayal16}. High-redshift galaxies have higher specific star formation rates \citep[e.g.][]{Feulner et al.} and a higher prevalence of active nuclei \citep[e.g.][]{Aird et al.}. Since high-redshift galaxies are also more compact \citep{Shibuya et al.}, supernovae, gamma-ray burst and active galactic nuclei will affect larger number of planets, and the size evolution may also increase the Oort cloud perturbations due to close encounters with other stars.

\begin{figure*}[t]
	\centering
	\includegraphics[width=84mm]{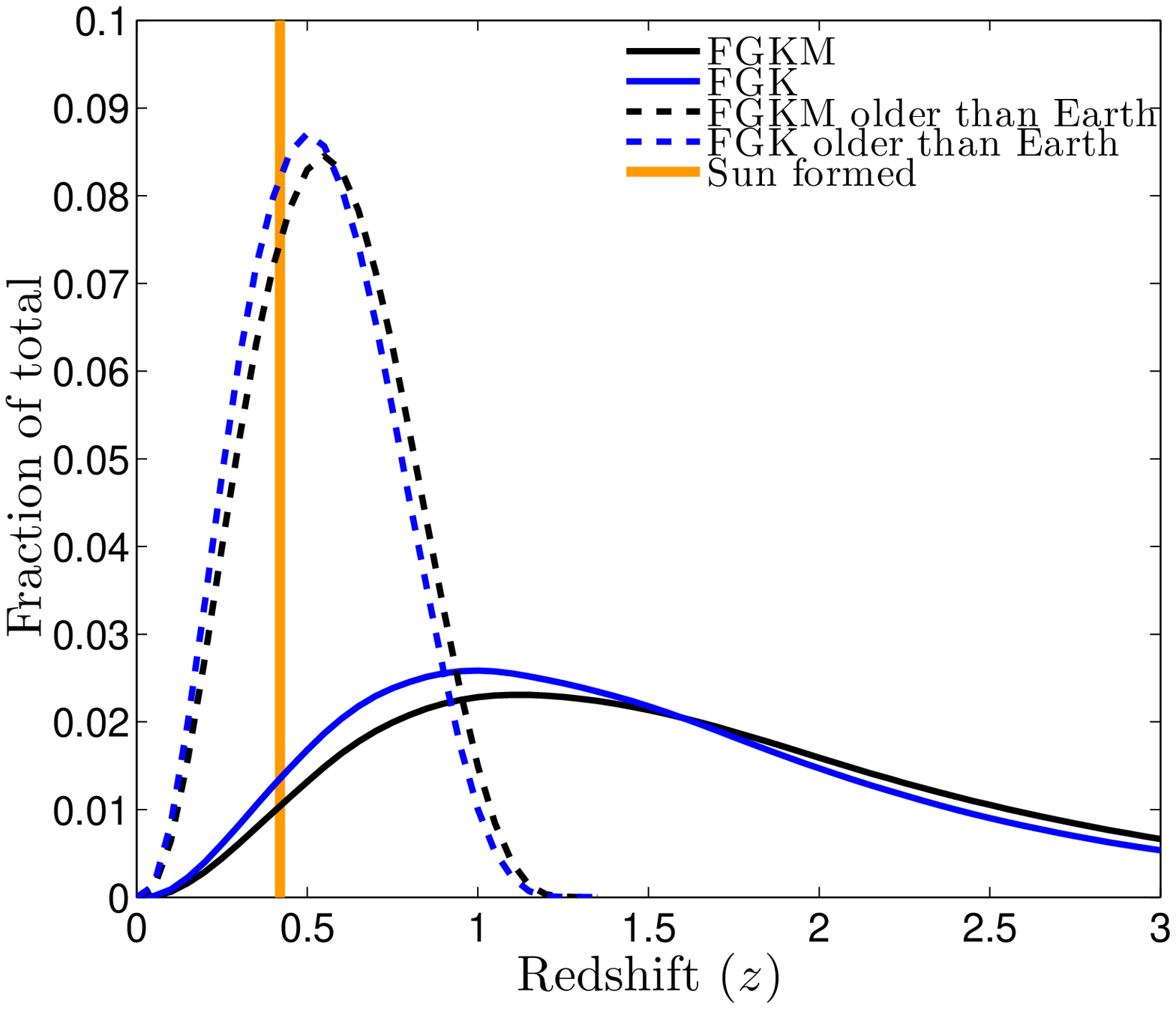}	
	\includegraphics[width=84mm]{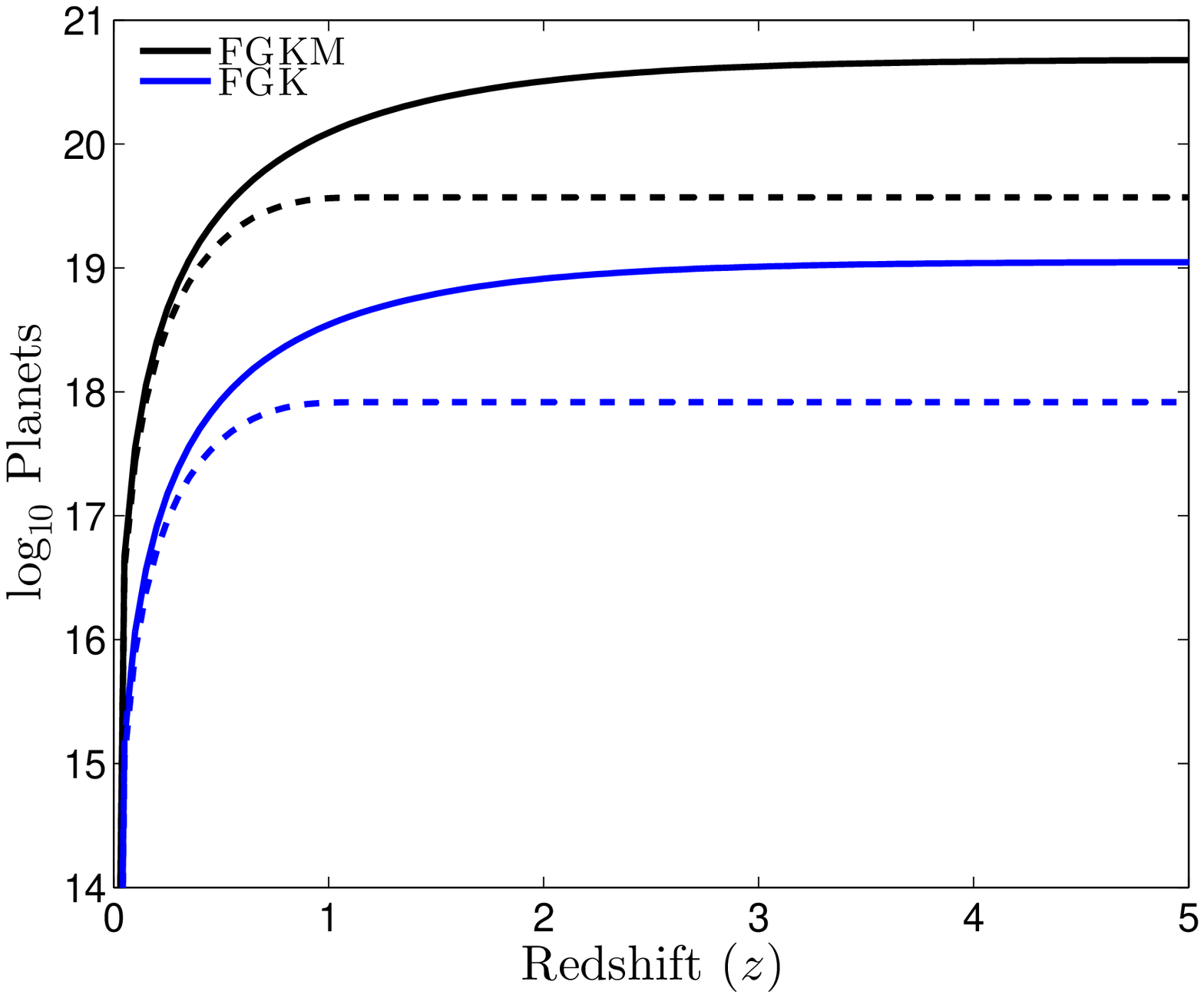}	
	\caption{Distribution of TPs as a function of redshift on our past light once. {\bf Left:} Distribution of TPs within the volume spanned by observable Universe. Solid lines mark the relative distribution of all FGKM (black line) and FGK (blue) host planets, and dashed lines the corresponding distributions when only planets older than Earth are considered. The orange patch marks the redshift that corresponds to the cosmic epoch when the Sun formed. {\bf Right:} The corresponding cumulative number of TPs in the whole sky out to the redshift given on the horizontal axis. At high redshifts, these quantities tend towards the numbers relevant for the whole observable Universe.}
	\label{planetsinsky}
\end{figure*}

\subsection{Why are we not living in an elliptical galaxy?}
\label{mediocrity_principle}
If we assume that all currently habitable terrestrial planets have a small but non-zero probability to develop intelligent life at around the present cosmic epoch, then the probability distribution for the emergence of intelligent life is expected to trace the spatial distribution of such planets. In other words, we expect to find ourselves in the type of environments where most such terrestrial planets are located. It is of course entirely possible that Earth could lie in the tail of this distribution, but this is an improbable outcome, and would violate the Copernican/mediocrity principle. 

By convolving the mass distribution of TPs from Figure~\ref{Planetdist_lowz} with the distribution of morphological types as a function of galaxy mass from \citet{Kelvin14} or \citet{Moffett16}, we estimate that $\approx 3/4$ of TPs are locked up in spheroid-dominated galaxies (with $\approx 1/2$ in elliptical galaxies and $\approx 1/4$ early-type disks), and not in disk-dominated galaxies like the Milky Way (likely morphological type Sbc). Hence, if Earth had been randomly drawn from the cosmological population of TPs, the most likely outcome would have been for it {\it not} to be located in a Milky Way-type system, but in an elliptical. While many spheroid-dominated galaxies do contain a disk component, and vice versa, TPs in disks remain in minority even when the estimated stellar mass fractions in bona fide disks and spheroids \citep{Moffett16b,Thanjavur16} are considered. 

This, by itself, does not constitute any significant violation of the Copernican/mediocrity principle, since there is nothing conspicuously improbable\footnote{At any rate no more improbable than the offsets between the properties of the Sun and other similar stars discussed by \citet{Robles et al.}}) about the type of host galaxy in which we happen to find ourself (probability $\approx 1/4$). However, the fact that the raw distribution of TPs is not already heavily skewed towards disk galaxies results in a slight puzzle once other habitability constraints are factored in, as these seem to tip the scales even further in favour of ellipticals.

When considering the role of supernova sterilization on galactic habitability, \citet{Dayal15} argue that massive ellipticals -- which exhibit much less current star formation activity than disks -- should exhibit much {\it better} conditions for life than actively star-forming galaxies like the Milky Way, and that this boosts the relative number of habitable planets per total stellar mass by 1-2 orders of magnitude in such galaxies at redshift $z\approx 0$. This implies that if our planet were randomly drawn from the cosmic distribution of habitable TPs at $z=0$, we should with $\gtrsim 90\%$ probability expect to find our self living in an elliptical galaxy.  

But maybe there is more to the lottery than we have hitherto realized? If supernovae are less detrimental for life than commonly assumed, or if some other life-threatening mechanism turns out to be more efficient in ellipticals than in disks, this could restore the balance between habitable TPs in spheroids and disks. We note that there are several important differences between disks and spheroids that have so far received very little attention in the context of astrobiology -- including the prevalence and properties of active galactic nuclei (but see \citealt{Dayal16} for a recent discussion on this), the propagation of cosmic rays and the fraction of stars in high-density regions. More detailed studies of the impact that mechanisms of this type can have on the habitability of planets may hopefully shed light on this matter.

\subsection{The prospects of extragalactic SETI}
\label{seti}
While most SETI projects have so far focused on the Milky Way, the case for Extragalactic SETI remains compelling \citep[e.g.][]{Wright et al. a,Zackrisson et al.,Olson}. If the probability for the emergence of intelligent life is sufficiently small, we could well be the only advanced civilization in the Milky Way. If so, our only chance of detecting intelligent life elsewhere in the Universe would be to extend the search radius. Extragalactic SETI would also seem the only way to gauge the prevalence of civilizations very high on the \citet{Kardashev} scale (close to type III), since the  Milky Way does not appear to host such a super-civilization.

If one adopts the admittedly anthropocentric view that intelligent life primarily develops on TPs, then our results have some bearing on the maximum distance to which it would make sense to push extragalactic SETI. Naively, one would expect that the probability for the emergence of an advanced civilization in an given cosmological volume is proportional to the number of TPs, modulo some correction for habitability and age effects. 

By extending the redshift limit of such searches, one increases the cosmological volume probed, but since one is -- due to the finite light travel time -- at the same time probing earlier epochs in the history of the Universe, the number of TPs within the search radius does not increase indefinitely. As discussed in section~\ref{highz} and shown in Figure~\ref{planetsinsky}b, the cumulative growth of the number of planets in the observable Universe slows down considerably beyond the peak of cosmic star formation ($z\approx 2$).  

If one furthermore assumes that it typically takes several billion years for intelligent life to arise (some 4.5 Gyr in the case of humans -- and please note that the \citet{Carter83} argument suggests that the typical time scale may be much longer than this), then only the $z\lesssim 2$ part of the Universe becomes relevant. For TPs older than Earth, only a very small fraction ($<10\%$) will be at $z>0.8$. When factoring in habitability considerations (section~\ref{habitability}), which will tend to further favour the low-density environment in the low-redshift Universe, and the escalating difficulties in detecting signatures (either signals or signs of astroengineering) of intelligent lifeforms at large distances, it seems that the prospects of extragalactic SETI should peak at redshifts below $z\approx 1$--2.

\subsection{Comparison to previous studies}
\label{comparison}
A small number of previous studies have attempted to assess the properties -- both ages and absolute numbers -- of terrestrial or Earth-like planets on cosmological scales. 

A coarse, early estimate of the total number of ``habitable planets'' (without clearly defining what was meant by this) in the observable Universe was presented by \citet{Wesson}. This estimate, despite being based on a different cosmological model and an unevolving galaxy population -- happens to lie within two orders of magnitude of our more modern estimate. By assuming $\sim 10^{10}$  habitable planets per galaxy and $\sim 10^{10}$ galaxies in the Universe, Wesson argues for $\sim 10^{20}$ habitable planets in the observable Universe. This is of the same order as our estimate on the total number of TPs (Table~\ref{statistics_highz}) and two orders of magnitudes higher than our estimate for the number of habitable planets around FGK stars (section~\ref{habitability}).

\citet{Lineweaver01} estimate the average TP in the $z=0$ Universe to be $1.8\pm 1.9$ Gyr older than Earth. While consistent with our estimate within the errorbars, this is still 1 Gyr lower than our best estimate for TPs around FGK stars ($\approx 2.8$ Gyr older than Earth; see Table~\ref{statistics_local}). 
As far as we can tell, this discrepancy is largely driven by the much stronger metallicity dependence adopted by Lineweaver, which pushes the peak of TP formation forward in time by cutting off planet formation at subsolar metallicities (see Figure \ref{planet_recipe} for a comparison). 

\citet{Behroozi & Peeples} also present formation time statistics for Earth-like planets (assumed to make up a fixed fraction of TPs around FGK stars) both in Milky Way-mass galaxies and over cosmological volumes. This is not expected to exactly match the age statistics at $z=0$ that we present for FGK host TPs in Table~\ref{statistics_local} and Figure~\ref{age_vs_mass_lowz}, since the current age distribution is affected both by the formation {\it and} demise of planets (due to the exhausted lifetimes of their host stars). However, since lifetime effects are unimportant for M dwarfs, we can still compare our results for M dwarf TPs with theirs to shed light on potential differences in the underlying computational machinery. In this case, we find that the difference in median ages is $<0.5$ Gyr. \citet{Behroozi & Peeples} also report that there are $\sim 10^{19}$ Earth-like planets on the past light cone, which -- when taking into account the scaling factor of $\approx 1/4$ between our mutual definitions of {\it terrestrial} and {\it Earth-like} -- lies within a factor of a few from our estimate.

A simple and commonly used exercise to estimate the number of terrestrial/habitable/Earth-like planets in the observable Universe is to adopt a literature value for the corresponding number in the Milky Way, assign this number to the large number of galaxies detected in a deep, high-redshift survey like the Hubble Ultra Deep Field or the Hubble Extreme Deep Field, and correct the outcome for the limited sky coverage of the survey. This estimate not only neglects differences in both metallicity and mass between high-redshift galaxies and the Milky Way, but also the completeness of the imaging survey used. Due to the redshift evolution of the galaxy mass function \citep[e.g.][]{Torrey15,Concelice16} and the redshift evolution of gas-phase metallicities in star-forming galaxies \citep[e.g.][]{Zahid13}, average galaxy masses and metallicties are both lower at high redshifts. The modest difference in the number of planets per stellar mass in the Milky Way compared to the cosmic average (cmp. Table \ref{statistics_local}  \& \ref{statistics_highz}) implies that mass, according to our models, is more important in this context. If we were to adopt our estimate for the number of TPs around FGKM stars in the Milky Way ($\approx 7\times 10^{10}$; see section~\ref{local}), multiply by the number of galaxies in the Hubble Ultra Deep Field ($\sim 10^4$) and correct for the limited sky coverage of the latter ($\approx 7.4\times 10^{-8}$), we would arrive at $\approx 9\times 10^{21}$ -- a factor of 20 higher than our estimate. The primary reason for this discrepancy is that the typical galaxy mass on our past light cone is {\it lower} than that of the Milky Way. In our simulations, the mean galaxy mass in the observable Universe (without any weighting by the probability of this object to host TPs) is $\approx 4\times 10^9\ M_\odot$. After applying this correction, this observational method results in an estimate that lies within a factor of $\approx 2$ of our result.

\subsection{Uncertainties}
\label{uncertainties}
A study such as this, which attempts to extrapolate results on the TP population from our local neighborhood to the whole Universe obviously carries substantial uncertainties. In the following, we attempt to identify the factors that dominate the error budget, and to attach reasonable error bars to our main results.

\subsubsection{Galaxy formation}
\label{uncertainties:galaxy formation}
The results presented in this paper rely heavily on semi-analytic models of galaxy formation to trace the formation of TPs across space and time, but consistency checks using observational data can in some cases be carried out to test the robustness of our results. For instance, the mean age of TPs is primarily determined by the cosmic star formation history, and while the star formation history predicted by GALFORM is already in reasonable agreement with observations at $z\lesssim 8$, we can derive an alternative estimate by simply adopting the fitting function for the cosmic star formation history presented by \citet{Madau & Dickinson} while keeping the metallicity distribution from GALFORM. The resulting average age differs from that presented in Table~\ref{statistics_local} by $\approx 0.2$ Gyr and in Table~\ref{statistics_highz} by $\approx 0.01$ Gyr. This approach also alters the total number of planets in the observable Universe by $\approx 20\%$. A comparison between our best estimate on the number of planets in the observable Universe and that resulting from either the \citet{Behroozi & Peeples} study, or from attaching the model-inferred number of planets per galaxy to galaxy counts in deep fields (section~\ref{comparison}), suggests consistency to within a factor of a few. We therefore adopt a conservative 0.5 dex uncertainty on all estimates of planet numbers in Tables~\ref{statistics_local} and \ref{statistics_highz}. We take the outcome of the minor age offsets seen when compared to the results of \citet{Behroozi & Peeples} in Section~\ref{comparison} to reflect the likely age uncertainties stemming from systematic uncertainties in semi-analytic models and conservatively adopt a 0.05 dex error on all age estimates. 

\subsubsection{Cosmological parameters}
Uncertainties on the parameters of the adopted cosmological model (within the framework of $\Lambda$CDM cosmology) primarily translate into an uncertainty on the age of the Universe as a function of redshift and on the calculation of cosmological volumes in our analysis, thereby shifting the age scale for the cosmic population of TPs and the planet counts. However, variations in $\Omega_\mathrm{M}$, $\Omega_\Lambda$ and $H_0$ at the level motivated by the differences between the parameter sets favored by the recent WMAP-9 \citep{Hinshaw13} and Planck \citep{Planck15} surveys have a relatively small impact on our results, amounting to age uncertainties of $<0.1$ Gyr and $\approx 0.2$ dex uncertainties in planet counts. A somewhat larger uncertainty comes from the fact that the galaxy catalogs originally generated by GALFORM \citep{Baugh et al.} were based on the $\Omega_{M}=0.3$, $\Omega_{\lambda}=0.7$, $H_0 = 70$ km s$^{-1}$ Mpc$^{-1}$ version of the $\Lambda$CDM cosmology, with $\sigma_{8}=0.93$ (a somewhat higher power spectrum normalization than favored by the \citealt{Planck15}), whereas the Galacticus catalogs were originally based on $\Omega_{M}=0.25$, $\Omega_{\lambda}=0.75$, $H_0 = 73$ km s$^{-1}$ Mpc$^{-1}$, $\sigma_{8}=0.9$. However, the shift in the age-redshift relation resulting from these slight inconsistencies remains at the $<0.3$ Gyr level. Other effects on the predicted galaxy population are degenerate with the many assumptions going into the semi-analytic machinery (e.g. chemical yields, the progenitors mass limits for different kinds of supernovae), giving errorbars on planet counts that are likely to be captured by the 0.5 dex uncertainty we argue for in section~\ref{uncertainties:galaxy formation}.

\subsubsection{Stellar initial mass function}
The results presented in this paper are based on the assumption that the fraction of FGKM stars can be estimated using the \citet{Kroupa01} universal stellar initial mass function (IMF). To assess how this affects our estimates of the number of TPs, we have recomputed the planet formation history assuming the \citet{Chabrier} instead of the \citet{Kroupa01} IMF when estimating the fraction of FGK and M stars forming within each age bin of the simulated galaxies. Both of these IMFs predict that $\approx 80\%$ of all stars in the 0.08--120 $M_\odot$ are born in the M dwarf mass regime (here 0.08--$0.6\ M_\odot$) whereas $\approx 10\%$ are born in the mass range of FGK stars (here 0.6--$1.2\ M_\odot$). Consequently, the predicted difference in TP statistics is very small -- at the 0.02 dex level for number statistics and 0.01 dex for age estimates. However, the possibility that the stellar IMF may be a function of both time and environment has been raised \citep[e.g.][]{Chabrier14}, and this could introduce uncertainties significantly larger than the ones considered here.

\subsubsection{Planet formation}
\label{Planet_formation_uncertainties}

The occurrence rates of TPs at low metallicities are admittedly highly uncertain, but as argued in Section ~\ref{local}, our results are remarkably robust to the reasonable adjustments of the planet occurrence recipe at the low metallicity end, simply because of the insignficant fraction of stars at such low metallicities. Adopting the minimum metallicity for the formation of low-mass planets (including the TPs discussed here) of [Fe/H] $\lesssim -1.8$ advocated by \citet{Hasegawa & Pudritz} would affect the number estimates at the sub-percent level. By assuming a step function similar to that of \citet{Prantzos} with a steep drop to zero probability for forming terrestrial planets at [Fe/H]$\leq 1.0$, we decrease the total number of terrestrial planets on our past light-cone by $6\%$. On the other hand, in a scenario where terrestrial planet formation has no metallicity dependency at all and we let the occurrence of terrestrial planets be constant at the low-metallicity end, we obtain an increment of terrestrial planets by $\approx 1\%$.

The possibility that TPs may reform in the wake of migrating Jupiters \citep[e.g.][]{Fogg & Nelson} may boost the fraction of TPs somewhat, but according to our estimates (Tables \ref{statistics_local} and \ref{statistics_highz}), this only affects the total population of TPs at the $\approx 10\%$ level for FGK hosts and at the $\approx 1$--2 \% level for M dwarf hosts. 

When it comes to the fraction of TPs in the circumstellar habitable zone, the uncertainties become much larger, since the observational occurrence rate of such planets around FGK stars alone is uncertain by almost an order of magnitude \citep[e.g.][]{Lissauer+14}, and the the habitability of both M-dwarfs and super-Earths are debatable.

\subsubsection{Total error budget}
Based on the considerations above, we adopt an error of $\approx 0.05$ dex on our age estimates (e.g. average TP age $8\pm 1$ Gyr for $z=0$ and $1.7\pm 0.2$ Gyr for our past light cone) and an error on the total number of TPs of 0.5 dex. This does not, however, include uncertainties related to potential variations of the shape/slope of the stellar initial mass function at the low-mass end with time and/or environment, since such variations remain poorly constrained and could have a very large impact on the number of M dwarfs.

\section{Summary}
\label{summary}
Our model for cosmic planet formation indicates that:
\begin{itemize}
\item The average age of TPs orbiting FGK stars in the local volume is $\approx 7\pm 1$ Gyr, whereas the corresponding value for TPs orbiting M dwarfs is $\approx 8\pm 1$ Gyr. The difference is primarily due to the limited lifetimes of FGK stars compared to the current age of the Universe. The average ages of TPs (around both FGK and M-type stars) in the whole observable Universe (on our past light cone) is 
$1.7\pm 0.2$ Gyr. The lower age compared to the local volume stems from the lookback-time effect, in which distant regions of the Universe are seen at an earlier cosmic epochs.

\item While the median TP in the local Universe is sitting in a galaxy of mass comparable to that of the Milky Way, the median TP on our past light cone is sitting in a galaxy with mass a factor of $\approx 5$ lower. 

\item The number of TPs per unit stellar mass at $z\approx 0$ remains almost constant (to within $\approx 30\%$) for galaxies in the stellar mass range $\sim 10^9$--$10^{11}\ M_\odot$. As a result, large satellite galaxies like the LMC are expected not to be depleted in TPs. Based on the observed mass-metallicity relation for dwarf galaxies and the recipe for metallicity-dependent TP formation adopted in this work, significant TP depletion is not expected to set in until $\lesssim 10^7\ M_\odot$, which is below the resolution limit of our simulations.

\item The total number of TPs in a Milky way-mass galaxy at $z=0$ is $\log_{10} N_\mathrm{TP}\approx 9.4\pm 0.5$ for TPs around FGK stars and $\log_{10} N_\mathrm{TP}\approx 10.8\pm 0.5$ for TPs around M dwarfs. 

\item The numbers of TPs in the whole observable Universe are $\log_{10} N_\mathrm{TP}\approx 19.0\pm 0.5$ (FGK hosts) and $\log_{10} N_\mathrm{TP}\approx 20.7\pm 0.5$ (M dwarf hosts).

\item There are conservatively $\sim 5\times 10^{-3}$ habitable TPs around FGK stars per unit total stellar population mass ($M_\odot$) and $\sim 10^{18}$ habitable TPs around FGK stars in the observable Universe

\item About $\approx 10\%$ of the TPs in the local Universe are orbiting stars at metallicities lower or higher than those for which such planets have so far been detected. The corresponding fraction for the whole observable Universe is estimated at $\approx 15\%$

\item Our model suggests that the typical TP in the local Universe is not sitting in a disk-dominated system like the Milky Way, but in a spheroid-dominated galaxy -- with about 1/2 of the  local TP population in ellipticals. This leads to a mild violation of the Copernican/mediocrity principle once habitability considerations related to supernova sterilization \citep{Dayal15} are factored in. Since the \citet{Dayal15} predicts much better conditions for life on planets in massive ellipticals compared to actively star-forming systems like the Milky Way, the distribution of habitable TPs would become biased in favour of ellipticals and make the Earth an outlier in the distribution of TP across the current galaxy population. 

\item Only a small fraction ($\leq 10\%$) of the TPs  on our past light cone are at redshifts $z\geq 3.3$, and when considering the subset of these planets older than Earth, this redshift limit drops to $z\gtrsim 0.8$. Under the assumption that the emergence of intelligent life is tied to the formation of TPs, the prospects of extragalactic SETI efforts are therefore expected to peak at relatively low redshifts. 

\end{itemize}

\acknowledgments
\vspace{5mm}
E.Z. acknowledge funding from the Magnus Bergvall foundation and Nordenskj\"{o}ldska Swendeborgsfonden. E.Z. and J.G. acknowledge funding from the Wenner-Gren Foundations. AJB acknowledges generous support from the Ahmanson Foundation, in the form of computing resources used for this work. A.J. and M.J. acknowledge funding from the Knut and Alice Wallenberg foundation.

\end{document}